\documentclass[twocolumn,amsmath,amssymb,floatfix]{revtex4}

\usepackage{latexsym,amsmath,amssymb,theorem,dsfont}
\usepackage{amssymb} 
\usepackage{amsmath}

\usepackage{amssymb}
\usepackage{amsmath}

\newcommand{\be}{\begin{equation}}
\newcommand{\ee}{\end{equation}}
\newcommand{\bea}{\begin{eqnarray}}
\newcommand{\eea}{\end{eqnarray}}
\newcommand{\beas}{\begin{eqnarray*}}
\newcommand{\eeas}{\end{eqnarray*}}

\newcommand{\half}{\frac{1}{2}}

\def\lsim{\mathrel{\mathstrut\smash{\ooalign{\raise2.5pt\hbox{$<$}\cr\lower2.5pt\hbox{$\sim$}}}}}
\def\gsim{\mathrel{\mathstrut\smash{\ooalign{\raise2.5pt\hbox{$>$}\cr\lower2.5pt\hbox{$\sim$}}}}}

\begin{document}

\title{Symmetron Fields: Screening Long-Range Forces Through Local Symmetry Restoration}

\author{Kurt Hinterbichler and Justin Khoury}

\affiliation{Center for Particle Cosmology, University of Pennsylvania, Philadelphia, PA 19104}
 
\begin{abstract}
We present a screening mechanism that allows a scalar field to mediate a
long range ($\sim$Mpc) force of gravitational strength in the cosmos while satisfying local tests of gravity.
The mechanism hinges on local symmetry restoration in the presence of matter. In regions of sufficiently high matter density, the field is drawn towards $\phi = 0$ where its coupling to matter vanishes and the $\phi\rightarrow -\phi$ symmetry is restored. In regions of low density, however, the symmetry is spontaneously broken, and the field couples to matter with gravitational strength. We predict deviations from general relativity in the solar system that are within reach of next-generation experiments, as well as astrophysically observable violations of the equivalence principle. The model can be distinguished experimentally from Brans-Dicke gravity, chameleon theories and brane-world modifications of gravity.

\end{abstract}

\maketitle 

Scalar fields are the simplest of fields. Light, gravitationally coupled scalars are generically predicted to exist by many theories of high energy physics.
These scalars may play a crucial role in dark energy as quintessence fields, and generically arise in infrared-modified gravity theories~\cite{DGP,cascading,galileon,f(R),Khoury:2003aq,Gubser:2004uf,ghost}. Despite their apparent theoretical ubiquity, no sign of such a fundamental scalar field has ever been seen, despite many experimental tests designed to detect solar system effects or fifth forces that would naively be expected if such scalars existed~\cite{Fischbach:1999bc,Will:2001mx}.  

Several broad classes of theoretical mechanisms have been developed to explain why such light scalars, if they exist, may not be visible to experiments performed near the Earth. One such class, the chameleon mechanism~\cite{Khoury:2003aq,Gubser:2004uf}, operates whenever the scalars are non-minimally coupled to matter in such a way that their effective mass depends on the local matter density. Deep in space, where the local mass density is low, the scalars would be light and would display their effects, but near the Earth, where experiments are performed, and where the local mass density is high, they would acquire a mass, making their effects short range and unobservable. 

Another such mechanism, the Vainshtein mechanism~\cite{Vainshtein:1972sx}, operates when the scalar has derivative self-couplings which become important near matter sources such as the Earth. The strong coupling near sources essentially cranks up the kinetic terms, which means, after canonical normalization, that the couplings to matter are weakened. Thus the scalar screens itself and becomes invisible to experiments. This mechanism is central to the phenomenological viability of brane-world modifications of
gravity~\cite{DGP,cascading} and galileon scalar theories~\cite{galileon}.

In this Letter, we explore a third class of mechanisms for hiding a scalar.
A similar framework was studied in~\cite{Olive:2007aj,Pietroni:2005pv} with different motivations, and some the results below overlap
with these works. In this mechanism, the vacuum expectation value (VEV) of the scalar
depends on the local mass density, becoming large in regions of low mass density, and small in regions of high mass density. 
In addition, the coupling of the scalar to matter is proportional to the VEV, so that the scalar couples with gravitational strength in regions
of low density, but is decoupled and screened in regions of high density.

This is achieved through the interplay of a symmetry-breaking potential, $V(\phi) = -\mu^2\phi^2/2 + \lambda\phi^4/4$, and universal coupling
to matter, $\phi^2 \rho/2M^2$. In vacuum, the scalar acquires a VEV $\phi_0 = \mu/\sqrt{\lambda}$, which spontaneously breaks
the $\mathds{Z}_2$ symmetry $\phi\rightarrow -\phi$. In the presence of sufficiently high ambient density, however, the field is confined near $\phi =0$, and the symmetry is
restored. In turn, $\delta\phi$ fluctuations couple to matter as $({\phi_{\rm VEV}}/M^2)\delta\phi\; \rho$, and so are
weakly coupled in high density backgrounds and strongly coupled in low density backgrounds. Since the screening mechanism relies on the local restoration of a symmetry, we refer to the scalar as a {\it symmetron} field.

The model predicts a host of observational signatures. The solar light-deflection and
time-delay deviations from general relativity (GR) are just below currents bound and within reach of
next-generation experiments. Meanwhile, the expected signal from binary pulsars is much weaker, because
neutron stars and their companions are screened. This is unlike standard Brans-Dicke (BD) theories, where solar
system and binary pulsar signals are comparable. The symmetron observables are similarly distinguishable from
standard chameleon and Vainshtein predictions. The symmetron also results in apparent violations 
of the equivalence principle between large (screened) galaxies and small (unscreened)
galaxies~\cite{Hui:2009kc}.

There are key differences with~\cite{Olive:2007aj,Pietroni:2005pv}, with crucial phenomenological implications. Because the symmetron is
universally coupled, we need not impose that the Earth and its atmosphere be screened, unlike~\cite{Olive:2007aj}.  
Instead, we show that a much weaker condition, namely that the Milky Way (and the Sun) be screened but not the Earth,
suffices to satisfy all local tests. This allows for longer range symmetron-mediated force. 
Compatibility with local tests of gravity was not considered in~\cite{Pietroni:2005pv}.

{\bf I. The Model:} Start with the general case of the chameleon model~\cite{Khoury:2003aq}, with metric signature $(-,+,+,+)$,
\bea
\nonumber
S&=&\int {\rm d}^4x\sqrt{-g}\left[ {M_{\rm Pl}^2\over 2} R-\half (\partial\phi)^2-V(\phi)\right]\\
&+& \int {\rm d}^4x\; {\cal L}_{\rm m}[\tilde{g}]\,,
\eea
where the matter fields described by ${\cal L}_{\rm m}$ are universally coupled to the metric $\tilde{g}_{\mu\nu}$,
conformally related to the Einstein frame metric $g_{\mu\nu}$ by
\be 
\tilde{g}_{\mu\nu} =  A^2(\phi) g_{\mu\nu}\,.
\label{conf}
\ee
The scalar field equation of motion is
\be
\square\phi-V_{,\phi}+A^3(\phi)A_{,\phi} \tilde{T}=0\,,
\ee
where $\tilde{T}= \tilde{T}_{\mu\nu}\tilde{g}^{\mu\nu}$ is the trace of the Jordan frame energy momentum tensor,
$\tilde{T}_{\mu\nu}= -(2/\sqrt{-\tilde{g}})\delta{\cal L}_{\rm m}/ \delta \tilde{g}^{\mu\nu}$, which is covariantly conserved: 
$\tilde{\nabla}_\mu  \tilde{T}^{\mu}_{\ \nu}=0$.

We will be interested mostly in solar system and galactic scenarios, so we ignore the effects of non-linearities in gravity and back-reaction of the scalar field on gravity, allowing us to treat the (non-linear) scalar on its own. 
For astrophysical objects, we may use the idealization of spherically symmetric pressureless sources.  Written in terms of the density $\rho=A^{3}\tilde\rho$, which is conserved in Einstein frame, the scalar field equation takes the form
\be 
{{\rm d}^2\over {\rm d}r^2}\phi+{2\over r}{{\rm d}\over {\rm d}r}\phi=V_{,\phi}+ A_{,\phi}\rho\,.
\label{sphericalequation} 
\ee
For cases of roughly homogeneous $\rho$, such as the interior or exterior of a star or galaxy, the field thus evolves according to an effective potential
\be 
V_{\rm eff}(\phi)=V(\phi)+\rho A(\phi)\,.
\ee

For the symmetron model of interest, we choose
\be 
V(\phi)=-{1\over 2}\mu^2\phi^2+{1\over 4}\lambda\phi^4, \ \ \ 
A(\phi)=1+{1\over 2M^2}\phi^2\,,
\label{ourpotential} 
\ee
described by two mass scales, $\mu$ and $M$, and one dimensionless coupling $\lambda$.  
The mass term in $V(\phi)$ is negative, so that the $\mathds{Z}_2$ symmetry $\phi\rightarrow -\phi$ is
spontaneously broken. The effective potential is, up to an irrelevant constant,
\be 
V_{\rm eff}(\phi)={1\over 2}\left({\rho\over M^2}-\mu^2\right)\phi^2+{1\over 4}\lambda\phi^4\,.
\ee
Whether the quadratic term is negative or not, and hence whether the $\mathds{Z}_2$ symmetry is spontaneously broken or not, depends on the local matter density.  Outside the source, where $\rho=0$, the potential breaks reflection symmetry spontaneously, and the scalar acquires a VEV
 $\phi_0\equiv \mu/\sqrt\lambda$.  Inside the source, if we choose parameters such that  $\rho > M^2\mu^2$, the effective potential no longer breaks the symmetry, and the VEV goes to zero.  
 
An essential feature is that the lowest order coupling of matter to the symmetron is $\sim\rho \phi^2/M^2$.  Fluctuations $\delta\phi$ around a local background value $\phi_{\rm VEV}$, as would be detected by local experiments, therefore couple as
\be 
\sim{\phi_{\rm VEV}\over M^2}\delta\phi \ \rho\,,
\label{coupling}
\ee
that is, the coupling is proportional to the local VEV.  In high-density, symmetry-restoring environments, the VEV should be near zero and fluctuations of $\phi$ should not couple to matter. In rarified environments, where $\rho < M^2\mu^2$, the symmetry is broken and the coupling turns back on.  

To fix scales, we will be mainly interested in the case where the field becomes tachyonic around the current cosmic density: $H_0^2M_{\rm Pl}^2 \sim \mu^2M^2$. This fixes $\mu$ in terms of $M$,
and hence the mass $m_0$ of small fluctuations around $\phi_0 = \mu/\sqrt{\lambda}$:
\be
m_0= \sqrt{2} \mu \sim \frac{M_{\rm Pl}}{M} H_0\,.
\label{muvalue}
\ee
Local tests of gravity, as we will see, require $M \lsim 10^{-3}M_{\rm Pl}$. Hence the range $m_0^{-1}$ of the symmetron-mediated force
in voids is $\lsim$~Mpc. Meanwhile, if this extra force is to be comparable to gravity, then from~(\ref{coupling}) we must impose
$\phi_0/M^2 \sim 1/M_{\rm Pl}$, that is,
\be
\phi_0\equiv \frac{\mu}{\sqrt{\lambda}} \sim \frac{M^2}{M_{\rm Pl}}\,.
\label{vev}
\ee
Together with~(\ref{muvalue}), this gives $\lambda \sim M_{\rm Pl}^4H_0^2/M^6\ll 1$. 
Note that $\phi_0 \ll M$, hence the field range of interest lies within the regime of the effective field theory,
and higher-order $\phi^2/M^2$ corrections to $A(\phi)$ can be neglected. 

{\bf II. Spherical solutions:} In order to model these effects in and around astrophysical objects, we search for spherically symmetric solutions to~(\ref{sphericalequation}) given~(\ref{ourpotential}). The object is taken to have radius $R$, and constant mass density $\rho$, such that $\rho > \mu^2M^2$. For simplicity, we further assume the object lies
in vacuum. Analogously to what is done in~\cite{Khoury:2003aq}, the radial field equation can be thought of as a fictional particle rolling in a potential $-V_{\rm eff}$,
subject to the ``friction'' term ${2\over r}{d\over dr}\phi$.  
The solution must be continuous at the origin, and approach its symmetry-breaking value far away from the object:
\be
\frac{{\rm d}}{{\rm d}r}\phi(0)=0\,;\ \ \ \ \phi(r\rightarrow \infty)=\phi_0\,.
\label{boundaryconditions}
\ee

We approximate the potential as quadratic around the appropriate minima both inside and outside the object, and then match at the surface of the object. 
Inside the object, we therefore have $V_{\rm eff}=\left({\rho\over M^2}-\mu^2\right)\phi^2/2$, and the solution satisfying the first of~(\ref{boundaryconditions}) is
\be 
\phi_{\rm in}(r)=A\, {R\over r}\sinh\left(r\sqrt{{\rho\over M^2}-\mu^2}\right)\,,
\label{insol}
\ee
with one undetermined constant $A$.  Outside the object, we approximate the potential as quadratic around the $\phi=\phi_0$ minimum:
$V_{\rm eff}=\mu^2(\phi-\phi_0)^2$. The solution satisfying the second of (\ref{boundaryconditions}) is
\be 
\phi_{\rm out}(r)=B\,{R\over r}e^{-\sqrt{2}\mu r}+\phi_0\,,
\label{outsol}
\ee
with one undetermined constant $B$.  Matching the field and its first derivative across the boundary at $r=R$ determines the two coefficients $A$ and $B$.  

The solutions involve three dimensionless parameters, $\mu R$, $\rho/\mu^2M^2$ and $\rho R^2/M^2$.
The first quantity measures the radius of the object relative to the range of the symmetron-mediated force in vacuum.
Since the latter is $\lsim$~Mpc, as seen earlier, for most objects of interest we have $\mu R\ll 1$. From~(\ref{muvalue}),
we recognize the second quantity as the density of the object as compared to the mean cosmic density. We will always
be interested in objects much denser than the cosmic mean, $\rho/\mu^2M^2\gg 1$.

Beyond that, we consider two cases depending on the value of $\rho R^2/ M^2$. Physically this ratio measures the surface Newtonian potential $\Phi$
relative to $M/M_{\rm Pl}$: 
\be
\alpha \equiv \frac{\rho R^2}{M^2} = 6\frac{M_{\rm Pl}^2}{M^2}\Phi \,.
\label{alp}
\ee
The first case is that of a small object, $\alpha \ll 1$. In this case we obtain
\be
A= \phi_0\frac{1}{\sqrt{\alpha}}  \left(1-{\alpha \over 2}\right)\,, \ \ B=-\phi_0{\alpha\over 3} \,.
\label{small}
\ee
The second case is that of a large object, $\alpha \gg 1$. 
\bea
A= \phi_0\frac{2}{\sqrt{\alpha}} e^{-\sqrt{ \alpha}}\,,\ \ 
B=\phi_0\left(-1 +\frac{1}{\sqrt{\alpha}}\right)\,.
\label{large}
\eea 

At distances $R \ll r\ll \mu^{-1}$, the force on a test particle due to a large object is suppressed compared to gravity:
\be
\frac{F_\phi}{F_{\rm N}} \sim \phi_0\frac{M_{\rm Pl}}{\rho R^2} = \frac{1}{\alpha} \ll 1\,,
\label{Ftest}
\ee
where we have used~(\ref{vev}) and~(\ref{alp}). This can be understood as an analogue of the thin-shell
effect of chameleon models. From~(\ref{insol}) and~(\ref{large}), we see that $\phi$ is exponentially suppressed compared to $\phi_0$
inside the object,  {\it e.g.} $\phi(r=0) = \phi_0 e^{-\sqrt{\alpha}}$, except for a thin shell of thickness
$\Delta R \sim \alpha^{-1}R$, within which $\phi\sim \phi_0/\sqrt{\alpha}$. The symmetron is
thus weakly coupled to the core of the object, and its exterior profile is dominated by the thin shell contribution.
In contrast, we see from~(\ref{insol}) and~(\ref{small}) that $\phi\approx\phi_0$ everywhere within
small objects. There is no thin shell in this case, and as a result $F_\phi/F_{\rm N} \sim {\cal O}(1)$.

{\bf III. Constraints from Tests of Gravity:} Since the force is long-range in all situations of interest, and because the symmetron couples to matter universally,
the tests to consider are the same that constrain standard BD theories: solar system and binary pulsar observations.

As we will see, to satisfy experimental constraints we will want the Milky Way to be screened: $\alpha_{\rm G} \gg 1$.
To get interesting cosmological effects, we focus on the limit where this condition is barely satisfied: $\alpha_{\rm G} \gsim 10$.
Since $\Phi \sim 10^{-6}$ for the galaxy, it follows from~(\ref{alp}) that
\be
M \lsim 10^{-3}M_{\rm Pl}\,,
\ee
as mentioned earlier. Using~(\ref{muvalue}), this implies that the symmetron-mediated force has $\lsim$~Mpc range in vacuum.
In this parameter regime, the Sun ($\Phi_\odot\sim 10^{-6}$) is screened, but the Earth ($\Phi_\oplus\sim 10^{-9}$) is not. 

What matters for solar system tests is the local field value, since this determines the coupling of the symmetron to matter.
At a generic point in the solar system, this is determined by the symmetron profile interior to the galaxy. Using~(\ref{insol}),~(\ref{large}) we find
\be
\frac{\phi_{\rm G}}{M} \approx \frac{M}{M_{\rm Pl}} \frac{R_{\rm G}}{\sqrt{\alpha_{\rm G}}R_{\rm us}} \exp\left(-\frac{R_{\rm G}-R_{\rm us}}{R_{\rm G}}\sqrt{\alpha_{\rm G}}\right)\,,
\ee
where $R_G\sim 100$~kpc is the Milky Way radius, and $R_{\rm us}\sim 10$~kpc is our distance from the galactic center. For $\alpha_{\rm G} = 20$, and
thus $M \approx 10^{-3}M_{\rm Pl}$, this gives $\phi_{\rm G}/M \approx 10^{-5}$.

\noindent {\it i) Time-delay and light-deflection observations}: 
The only non-trivial post-Newtonian parameter in this case is $\gamma$, defined in Jordan frame by $\tilde{g}_{00} = -(1+2\Phi_{\rm J})$ and $\tilde{g}_{ij} = 1 - 2\Phi_{\rm J}\gamma$, where $\Phi_{\rm J}$ is the Jordan-frame gravitational potential. Starting in Einstein frame, we have checked that the backreaction of $\phi$ on the metric is
negligible, {\it i.e.} $g_{00} \approx  -(1+2\Phi_{\rm E})$ and $g_{ij} \approx 1 - 2\Phi_{\rm E}$. Then, using~(\ref{conf}) to
translate to Jordan frame, we therefore have
\be
\gamma - 1 \approx - \frac{\phi^2}{M^2\Phi}\,.
\label{gamdef}
\ee
At this order $\Phi$ can be calculated in either frame. 

The signal for time-delay and light-deflection experiments is due primarily to photons passing near the Sun's surface ($d\sim R_\odot$).
Hence we need the value $\phi_\odot$ in the vicinity of the Sun. The Sun is screened,
therefore~(\ref{outsol}) and~(\ref{large}) apply, except with $\phi_0$ replaced by $\phi_{\rm G}$
to account for the galactic background density. We therefore obtain $\phi_\odot \approx \phi_{\rm G}/\sqrt{\alpha_\odot}$.
Substituting into~(\ref{gamdef}), the current constraint of $|\gamma - 1| \approx 10^{-5}$ from the Casini spacecraft~\cite{Bertotti:2003rm}
implies $\phi_{\rm G}/M \lsim 10^{-11/2}\alpha_{\rm G}$. Our fiducial choice $\alpha_{\rm G} = 20$ ($\phi_{\rm G}/M \approx 10^{-5}$) barely satisfies this bound,
thus our predicted signal is just below current sensitivity levels.

\noindent {\it ii) Nordvedt Effect:} This constrains the difference in free-fall acceleration between the Moon and the Earth towards the Sun, arising
from $\phi$-corrections to their self-gravity. It is easy to show that the change in the Earth's gravitational self-energy is
$|\Delta E_g/E_g|\sim (\phi(R_\oplus)-\phi_{\rm G})^2/M^2\Phi_\oplus$. A similar expression for the Moon yields a negligible contribution.
Since the Earth is unscreened, we can use~(\ref{small}) to predict the Nordvedt parameter $\eta_{\rm N}$
\be
|\eta_{\rm N}| \sim  \frac{(\phi(R_\oplus)-\phi_{\rm G})^2}{M^2\Phi_\oplus} = \frac{\alpha_\oplus^2}{4\Phi_\oplus}\left(\frac{\phi_{\rm G}}{M}\right)^2\,.
\ee
Since $\alpha_\oplus \approx 10^{-3}\alpha_{\rm G}$ and  $\Phi_\oplus = 10^{-9}$, the current bound $|\eta_{\rm N}| \lsim 10^{-4}$
from Lunar Laser Ranging observations~\cite{Will:2001mx} implies $\alpha_{\rm G}\phi_{\rm G}/M \lsim 2\times 10^{-7/2}$. This is barely satisfied
for our fiducial $\alpha_{\rm G} = 20$, and thus the symmetron signal is within reach of next-generation experiment~\cite{apollo}.

\noindent {\it iii) Perihelion Shift of Mercury:} Near Mercury, the field profile due to the Sun is $\phi(r) \approx \phi_{\rm G} (1-R_\odot/r)$. 
Using this, we find $|\gamma - 1| \approx \phi_{\rm G}(\phi -\phi_{\rm G})/M^2|\Phi_\odot|$. The current limit $|\gamma - 1|\lsim 10^{-3}$~\cite{Will:2001mx}
therefore implies $\phi_{\rm G}/M \lsim 10^{-9/2}$, which is satisfied for our fiducial $\phi_{\rm G}/M \approx 10^{-5}$.

\noindent {\it iv) Binary Pulsars:} Constraints from binary pulsars are trivially satisfied in our scenario, since both the neutron star
and its companion are screened:
\be
F_\phi = \alpha_{\rm pulsar}^{-1}\alpha_{\rm companion}^{-1} F_{\rm N}\,.
\ee
Estimating $\Phi_{\rm pulsar} \sim 0.1$ and $\Phi_{\rm companion}\sim 10^{-6}$, we obtain $F_\phi/F_{\rm N} = 10^{-5}/\alpha_{\rm G}^2$. 
The current constraint on BD theories translates to $F_\phi/F_{\rm N}\approx 1/2\omega_{\rm BD} \lsim 5\times 10^{-4}$~\cite{Will:2001mx}, which in our case
implies $\alpha_{\rm G}^2 \gsim 2\times 10^{-2}$. This is automatically satisfied by our earlier requirement $\alpha_{\rm G}\gsim 10$.

{\bf IV. Observational Signatures:} As we have seen above, our model predicts deviations from GR in the solar system that are comparable to current constraints and
therefore within reach of future experiments. Moreover, unlike standard BD theories where all predictions are determined by a single
parameter, here different observables correspond to different values of $\omega_{\rm BD}^{\rm eff}$. Most strikingly, the
predicted signal for binary pulsars is far weaker than for solar system tests.

The symmetron is also distinguishable from other screening mechanisms. In standard chameleon theory, the tightest constraint comes
from laboratory tests of the inverse square law. This results in unobservably small ($\omega_{\rm BD}^{\rm eff} \gsim  10^{12}$)
signals for solar system tests. In the Vainshtein case, brane-world gravity theories~\cite{DGP,cascading} yield modifications to
the Moon's orbit that are accessible to the next generation lunar ranging experiment~\cite{DGZ,degrav}, but the light deflection and
time delay signals are negligible.

The symmetron also results in apparent violations of the equivalence principle between large
(screened) and small (unscreened) galaxies, which can show up in various
astrophysical observations~\cite{Hui:2009kc}. For the fiducial parameters considered here, the threshold
gravitational potential below which objects are unscreened is $\Phi \sim 10^{-7}$. Typical dwarf galaxies are therefore unscreened.

{\bf V. Quantum Corrections:} The symmetron model is a more natural-looking effective theory than chameleon/$f(R)$ models, which typically involve exponential and inverse power potentials containing an infinite number of non-renormalizable operators. The symmetron has the $\mathds{Z}_2$ symmetry $\phi\rightarrow -\phi$, and its self-interactions are the most general renormalizable terms consistent with this symmetry. The quadratic coupling to the matter stress tensor is the leading such coupling compatible with the $\mathds{Z}_2$ symmetry.  It is non-renormalizable, suppressed by the mass scale $M$, thus we treat the symmetron as an effective theory with cutoff $\Lambda\sim M$.

As usual with conformally-coupled scalar fields, the symmetron potential receives large quantum corrections from matter loops. In particular, the symmetron mass gets $\sim m_{\rm SM}^2/M$ renormalization from Standard Model fields. Our analysis relies on these contributions being fine-tuned away, as in any scalar field model without shift symmetry. 

%Naively the symmetron mass receives large contributions $\sim m_{\rm SM}^2/M$ from Standard Model fields. Because the symmetron couples conformally, the result has identical
%coefficient to the vacuum energy contribution and hence is automatically set to zero by the usual fine-tuning of the cosmological constant~\cite{longpaper}. The scalar mass also receives a contribution from the $\lambda\phi^4$ self interaction vertex, $\delta \mu^2\sim \lambda \mu^2$, which is $\ll \mu^2$. These conclusions rely on dimensional regularization, which makes the most optimistic assumptions about the UV completion. 

{\it Acknowledgments:} We thank L.~Hui, H.~Nastase, A.~Nicolis, M.~Pospelov, D.~Stojkovic, M.~Trodden and D.~Wesley for
helpful discussions. This work was supported by funds provided by the University of Pennsylvania and
by NSERC of Canada.

\end{document}